\documentclass[10pt,conference]{IEEEtran}
\IEEEoverridecommandlockouts

\usepackage{geometry}
\usepackage{amsmath,amssymb,amsfonts}
\usepackage{algorithmic}
\usepackage{graphicx}
\usepackage{textcomp}
\usepackage{xcolor}
\usepackage{booktabs}
\usepackage{breqn}
\usepackage{tikz}
\usepackage{amsthm}
\usepackage{amsmath}
\usepackage[linesnumbered,ruled]{algorithm2e}
\usepackage{comment}
\usepackage{pst-all}
\usepackage{nicematrix}
\usepackage{multirow}
\usepackage{nomencl}
\usepackage{cite}

\newcommand{\sub}[1]{{\color{red} [Sub: #1]}}
\newcommand{\mil}[1]{{\color{green!30!black} [Mil: #1]}}
\newcommand{\bis}[1]{{\color{yellow!30!black} [Bis: #1]}}

\renewcommand\nomgroup[1]{%
  \item[\bfseries
  \ifstrequal{#1}{S}{Indices and Sets}{%
  \ifstrequal{#1}{P}{Parameters}{%
  \ifstrequal{#1}{V}{Variables}{}}}%
]}

\makenomenclature

\begin{document}

\title{
A Distributed Local Energy Market Clearing Framework Using a Two-Loop ADMM Method
\vspace{-3mm}}
\author{\IEEEauthorblockN{Milad Kabirifar\IEEEauthorrefmark{1}, Biswarup Mukherjee\IEEEauthorrefmark{2}, S. Gokul Krishnan\IEEEauthorrefmark{2},\\ Charalambos Konstantinou\IEEEauthorrefmark{2}, Subhash Lakshminarayana\IEEEauthorrefmark{1}
}
\IEEEauthorblockA{
\IEEEauthorrefmark{1}School of Engineering, University of Warwick, Coventry, United Kingdom \\
\IEEEauthorrefmark{2}CEMSE Division, King Abdullah University of Science and Technology (KAUST), Saudi Arabia\\
}

\vspace{-3em}} 

\maketitle

\begin{abstract}
The diversity of prosumers’ resources in energy communities can provide significant technical and economic benefits to both prosumers and the distribution system operator (DSO). To maximize these benefits, a coordination framework is required to address all techno-economic constraints as well as the objectives of all agents. This paper presents a fully distributed market-clearing scheme to coordinate the strategies of agents within a local energy community. In the proposed framework, prosumers, the DSO, and the local market operator (LMO) are the participating agents. The framework addresses the preferences and techno-economic constraints of all actors while preserving their privacy. The proposed model is based on a modified alternating direction method of multipliers (ADMM) method with two outer and inner loops; the outer loop models the interactions between the LMO and prosumers, while the inner loop addresses the interactions between the LMO and the DSO. The model is demonstrated on IEEE-69bus test network, showcasing its effectiveness from various perspectives.
\end{abstract}
\vspace{-1.5mm}
\bibliographystyle{IEEEtran}

\nomenclature[S]{$t,\tau$}{Hourly time intervals.}
\nomenclature[S]{$f,(n,m)$}{Distribution network lines (nodes).}
\nomenclature[S]{$a$}{Prosumers/aggregators participating in LEM.}
\nomenclature[S]{$u, b, e, l$}{PV units, BESSs, EVs, and FLs of prosumers.}
\nomenclature[S]{$v$}{Prosumers’ resources: PVs, BESSs, EVs, and FLs.}
\nomenclature[S]{${{\Omega }^N}$}{Set of network nodes including ${\Omega }_n^N$, ${\Omega }_a^N$, ${\Omega }^{N,{\text{PCC}}}$ indicating connected nodes to node $n$, nodes hosting agent $a$’s resources, and substation nodes.}
\nomenclature[S]{${{\Omega }^T}$}{Set of time intervals.}
\nomenclature[S]{${{\Omega }^F}$}{Set of distribution network feeders.}
\nomenclature[S]{${{\Omega }^A}$}{Set of prosumers participating in the LEM.}
\nomenclature[S]{${\Omega }_n^A$}{Set of participating agents located in bus $n$.}
\nomenclature[S]{${\Omega }_a^U, {\Omega }_a^B, {\Omega }_a^E, {\Omega }_a^L$}{Set of PV units, BESSs, EVs, and FLs.}
\nomenclature[S]{${\Omega_a ^V}$}{Set of prosumers' resources.}

\nomenclature[P]{\(r_{n,m},\,x_{n,m}\)}{Network lines' resistance and reactance.}
\nomenclature[P]{$\overline S_{n,m}^F$}{Capacity of distribution network lines.}
\nomenclature[P]{${\underline V _n},{\overline V_n}$}{Minimum and maximum bus voltage magnitudes.}
\nomenclature[P]{\(P_{a,u,t}^{\text{PV}}\)}{Forecasted active power of PV units.}
\nomenclature[P]{\(\overline S_{a,u}^{\text{PV}}, PF_{a,u,t}^{\text{PV}}\)}{Capacity and power factor of PV inverters.}
\nomenclature[P]{$\overline P _{a,b/e}^{\text{Ch-ESS/EV}},\,\,\overline P _{a,b/e}^{\text{dCh-ESS/EV}}$}{BESS/EVs' charge-discharge limits.}
\nomenclature[P]{$E_{a,e}^{\text{trip-EV}}$}{Required energy for EV's next trip.}
\nomenclature[P]{$\eta _{a,b/e}^{\text{Ch-ESS/EV}},\eta _{a,b/e}^{\text{dCh-ESS/EV}}$}{Charging and discharging efficiencies of BESSs/EVs.}
\nomenclature[P]{$E_{a,b/e}^{\text{0-ESS/EV}}$}{Initial charge of BESSs/EVs.}
\nomenclature[P]{\(\underline {SoC} _{a,b}^{\text{ESS}},\overline {SoC} _{a,b}^{\text{ESS}}\)}{Allowable limits of BESSs’ SoC.}
\nomenclature[P]{$a_{a,e}^{\text{EV}},\,d_{a,e}^{\text{EV}}$}{Arrival and departure times of EVs.}
\nomenclature[P]{$S_{a,e}^{\text{EV}}$}{Capacity of EVs' batteries.}
\nomenclature[P]{$\overline P_{a,t}^{\text{FL}},\underline E _a^{\text{L}}$}{Maximum amount of prosumers’ flexible load and minimum required energy of prosumers.}
\nomenclature[P]{$T_a^{\text{FL}}$}{Agreement time for modifying prosumers’ load.}
\nomenclature[P]{\(P_{a,t}^{\text{L}}\)}{Prosumers’ load before load interruption.}
\nomenclature[P]{\(\lambda _{t}^{\text{WEM}}, \Delta t\)}{WEM price and duration of time interval.}
\nomenclature[P]{$\Psi _{n,a}$}{Matrix indicating the location of prosumers at the distribution network.}

\nomenclature[V]{${\upsilon _{n,t}},\ell _{n,m,t}$}{Buses’ voltage \& square of feeders’ current.}
\nomenclature[V]{\(p_{n,m,t}^{\text{F}},q_{n,m,t}^{\text{F}}\)}{Active and reactive power flow of distribution network lines.}
\nomenclature[V]{\(p_{n,t}^{\text{net}},\,q_{n,t}^{\text{net}}\)}{Net active and reactive powers at each node $n$.}
\nomenclature[V]{\(p_{t}^{\text{UG}},q_{t}^{\text{UG}}\)}{Upstream grid active and reactive powers.}
\nomenclature[V]{\(p_t^{\text{Loss}}\)}{Total power losses of distribution network.}
\nomenclature[V]{\(\lambda _{n,t}^{\text{LEM}}\)}{Local energy market price (DLMP).}
\nomenclature[V]{\(p_{a,u,t}^{\text{PV}},q_{a,u,t}^{\text{PV}}\)}{Active and reactive power of PV units.}
\nomenclature[V]{$x_{a,b/e,t}^{\text{Ch-ESS/EV}},x_{a,b/e,t}^{\text{dCh-ESS/EV}}$}{Binary variables indicating charging and discharging status of BESSs/EVs.}
\nomenclature[V]{\(p_{a,b/e,t}^{\text{Ch-ESS/EV}},p_{a,b/e,t}^{\text{dCh-ESS/EV}}\)}{Charging and discharging power of BESSs/EVs.}
\nomenclature[V]{$SoC_{a,b/e,t}^{\text{ESS/EV}}$}{BESSs/EVs state of charge.}
\nomenclature[V]{$y_{a,t}^{\text{FL}}$}{Binary variable of FLs' utilization status.}
\nomenclature[V]{\(p_{a,t}^{\text{FL}}\)}{Change in the consumption power of prosumers.}
\nomenclature[V]{\(p_{a,t}^{\text{G}},p_{a,t}^{\text{L}}\)}{Generation and consumption power of prosumers.}
\nomenclature[V]{\(p_{a,t}^{\text{net}}\)}{Net consumption power of prosumers.}
\nomenclature[V]{\(c_{v,t}^{V}\)}{Operation cost of resource $v$.}

\printnomenclature[4em]

\section{Introduction}\label{intro}
With the transition from passive consumers to active prosumers in smart distribution networks, various prosumer resources can be leveraged to benefit both the prosumers and the underlying distribution network \cite{r1-ghasemnejad2024energy}. To optimize the use of these resources, a coordination framework is necessary \cite{r3-hu2020coordinated}. Local energy market (LEM) frameworks within distribution networks can effectively coordinate the strategies of agents in local energy communities \cite{r4-cai2016self}. Several LEM frameworks and clearing methodologies have been developed in the literature. In \cite{dlmp5-papalexopoulos2020development,24a1-bai2017distribution,30a1-yuan2016distribution}, a centralized market-clearing approach is presented, where a central agent manages the resources within the LEM to coordinate their operation. However, these methods do not model the preferences of the agents, such as prosumers, and fail to preserve the privacy of prosumers. 
To establish the hierarchical relationship of different agents by addressing their aims and constraints in the LEM clearing process, the authors in \cite{Integ1-chen2021distribution,Integ2-chen2021local,pourghaderi2020energy} propose an integrated mathematical optimization framework to model the LEM structure. While the model considers the agents’ preferences, the solution method for LEM clearing relies on a centralized approach that integrates all agents' models and data into a single optimization problem. 

To address the aforementioned challenges, a distributed optimization methodology is required for clearing the LEM overseeing local energy communities. 
A crucial signal used to clear the LEM with multiple actors is the distribution locational marginal pricing (DLMP) which reflects the energy price at each bus of the distribution network \cite{dlmp6-li2021electricity}. In \cite{distri-jacquot2020dlmp}, the coordination of demand response potentials in distribution networks is explored using DLMP as a reference signal. 
In \cite{admm2-ullah2022dlmp}, a two-level model is developed to represent the transactive energy trading of prosumers. In the upper level, a cost minimization problem is solved using the alternating direction method of multipliers (ADMM) method. In the lower-level problem, a non-cooperative peer-to-peer energy pricing game is formulated to determine the equilibrium price. In \cite{dlmp6-li2021electricity}, a linearized ACOPF model is developed to calculate DLMP and provide price signals for distributed generators (DGs) and aggregators within the distribution system. First, the DSO calculates the DLMP, and then aggregators trade with each other based on cooperative game theory. In these studies, the trading among different aggregators is modeled using game theory approaches; however, their interactions with the DSO and LMO through the LEM are not thoroughly explored\textcolor{black}{, which could lead to network constraint violations}.

In \cite{17a1-morstyn2018bilateral,18a1-xiao2017local}, a distributed framework for LEM clearance is presented; however, the impact of the physical distribution network on the pricing mechanism is neglected.
In \cite{admm1-he2021multi}, a distribution market pricing mechanism for utilizing the demand response of active customers is presented. To calculate the DLMP, the market clearing model is formulated as a multi-period ACOPF. The augmented Lagrangian relaxation method is employed to solve this problem. While the proposed method is valuable, its convergence to the optimal solution cannot be guaranteed when limiting the transacted data to preserve customers’ privacy. In \cite{revision-franke2024privacy} and \cite{Revision2-erdayandi2024pp}, privacy-aware methods are proposed for local market clearing. However, the emphasis is on developing secure market protocols, while the modeling of agent interactions and the operational constraints of the local energy market is not addressed. \textcolor{black}{In summary, the literature lacks an efficient distributed framework to coordinate the various actors within a local energy community, addressing the network's constraints and the preferences and detailed technical constraints of all agents while preserving their privacy.} 

To fill-up this gap, this paper presents a \textcolor{black}{distributed} energy trading mechanism based on precise DLMP calculations to coordinate prosumers' strategies within local energy communities through the LEM framework. In the proposed model, the LEM is cleared by the LMO, a non-profit agent, in collaboration with the DSO and prosumers. The DSO, acting as a technical agent, addresses the operational constraints of the distribution network. The DLMP is calculated through the collaboration between the LMO and DSO using a second-order cone programming (SOCP) model. Prosumers then optimize their strategies to maximize profit by participating in the LEM. The proposed \textcolor{black}{distributed} LEM clearing formulation is solved using a modified ADMM method with two loops: an outer loop that establishes the interaction between the LMO and prosumers, and an inner loop that coordinates the strategies of the LMO and DSO. \textcolor{black}{The proposed model ensures that the privacy of all agents will be preserved.}
\vspace{-5 pt}
\section{Proposed Framework}\label{framework}
\vspace{-5 pt}
\subsection{Methodology}\label{method}
\vspace{-1mm}
This paper presents a market-based framework to coordinate the strategies of prosumers in a local energy community. In this framework, prosumers participate in the day-ahead (DA) LEM to determine the optimal schedule for their resources, aiming to maximize their profits. The LMO acts as a non-profit entity responsible for clearing the LEM to maximize social welfare. The LMO can participate in the wholesale energy market (WEM). \textcolor{black}{In the proposed model, the LMO receives electricity price from the WEM and optimizes its strategy based on this signal}. To develop a feasible strategy, the impact of distribution network must be considered. Accordingly, the LMO interacts with the DSO during the LEM clearance. As a technical agent, the DSO addresses the operational constraints of the network and interacts with the LMO, aiming at minimizing energy losses.
The prosumers’ resources include photovoltaic (PV) units, BESSs, EVs, and flexible loads (FLs). 
It is worth noting that the proposed model is general and can similarly accommodate the participation of aggregators and DG owners.

The proposed DA-LEM is cleared using a fully distributed solution methodology, addressing agent interactions as well as all agents' preferences and techno-economic constraints. The transaction signals between LEM actors are designed to preserve agents' privacy, ensuring that internal confidential data does not need to be exchanged during the LEM clearance procedure. \textcolor{black}{In this vein, the LMO needs to transact the DLMP of the buses with the corresponding prosumers, while the prosumers share their net consumption power with the LMO. In addition, the DSO shares the total network losses with the LMO, while the LMO sends the net consumption power at each bus to the DSO.} The method utilizes a two-loop modified ADMM with convex sub-problems, which is a mathematically efficient approach for solving distributed optimization problems. The LEM is cleared by modeling the DLMP, which specifies the energy price at each bus in the distribution network. The DLMP reflects the impacts of feeders’ congestion, losses, and voltage support on prices.
\vspace{-5 pt}
\subsection{Mathematical Formulation}\label{math}
\subsubsection{DSO Model}
The DSO is responsible for operating the distribution network to ensure the secure and reliable supply of demand. The DSO collaborates with the LMO to ensure that all network operational constraints are satisfied during LEM clearance. In summary, the DSO monitors local market transactions to ensure feasibility and manages the network to achieve its objectives. The DSO aims to minimize total network losses, as specified in \eqref{dso_ob}.
\vspace{-5 pt}
\begin{equation}
    \textbf{Min}\,\,\sum\limits_{t \in {\Omega }^T} {p_t^{\text{Loss}}.C_t^{\text{Loss}}.\Delta t} \
    \label{dso_ob}
\end{equation}
The active and reactive power balance must be maintained at each bus in the distribution network. Sets of equations \eqref{active_bal} represents the active power balance and the reactive power balance can be formulated similarly. $\lambda _{n,t}^{\text{LEM}}$ is the dual variable associated with \eqref{active_bal} which represents the DLMP of LEM.  
\begin{equation}
\begin{array}{l}
\sum\limits_{m \in {\Omega }_n^N} {p_{n,m,t}^{\text{F}} + p_{n,t}^{\text{net}}\, = \,} p_{t}^{\text{UG}}\left| {_{n \in{\Omega }^{N,{\text{PCC}}} }} \right.\text{:}\,\,\lambda _{n,t}^{\text{LEM}},\\
\forall \,n\, \in {{\Omega }^N},\,t\, \in {{\Omega }^T}\
\end{array}\
\label{active_bal}
\end{equation}

To model the distribution network load flow equations and calculate the DLMP, SOCP model is employed. SOCP is a relaxed form of semi-definite programming (SDP) model and serves as an effective model for capturing the impacts of line congestion, power losses, and voltage support in the DLMP calculation, while accounting for the coupling of active and reactive power in the network \cite{30a1-yuan2016distribution}. The branch flow model is used to describe the network's operational model \eqref{ldf} – \eqref{volt_mag}.
\begin{equation}
\begin{array}{l}
\upsilon _{n,t} - \upsilon _{m,t} - 2\left( {r_{n,m}.p_{n,m,t}^{\text{F}} + x_{n,m}.q_{n,m,t}^{\text{F}}} \right) + \\
\ell _{n,m,t}\left( {r_{n,m}^2 + x_{n,m}^2} \right) = 0\text,
\forall \,n\,\in{{\Omega }^N},\,m\,\in {\Omega }_n^N,\,t\,\in{{\Omega }^T}
\end{array}\
    \label{ldf}
\end{equation}
\begin{equation}
\begin{array}{l}
{\left( {p_{n,m,t}^{\text{F}}} \right)^2} + {\left( {q_{n,m,t}^{\text{F}}} \right)^2} \le {\upsilon _{n,t}}\ell _{n,m,t},\\
\forall \,\,n\, \in {{\Omega }^N},\,m\, \in {\Omega }_n^N,\,t\, \in {{\Omega }^T}
\end{array}
    \label{ldf2}
\end{equation}
\vspace{-10 pt}
\begin{equation}
\begin{array}{l}
{\left( {p_{n,m,t}^{\text{F}}} \right)^2} + {\left( {q_{n,m,t}^{\text{F}}} \right)^2} \le {\left( {\overline S_{n,m}^F} \right)^2},\\
\forall \,\,n\, \in {{\Omega }^N},\,m\, \in {\Omega }_n^N,\,t\, \in {{\Omega }^T}
\end{array}    
\label{line_cap}
\end{equation}
\begin{equation}
\begin{array}{l}
{\left( {{{\underline V }_n}} \right)^2} \le {\upsilon _{n,t}} \le {\left( {{{\overline V}_n}} \right)^2},
\forall \,\,n\, \in {{\Omega }^N},\,t\, \in {{\Omega }^T}
\end{array}
\label{volt_mag}
\end{equation}
The exact form of \eqref{ldf2} is ${\left( {p_{n,m,t}^{\text{F}}} \right)^2} + {\left( {q_{n,m,t}^{\text{F}}} \right)^2} = {\upsilon _{n,t}}\ell _{n,m,t}$  which makes the problem non-convex and it is relaxed as \eqref{ldf2} to transform the problem into a SOCP problem. The total loss of the grid at each time interval is determined in \eqref{loss}.
\begin{equation}
    p_t^{\text{Loss}} = \sum\limits_{\left( {n,m} \right) \in {{\Omega }^F}} {{r_{n,m}}.\ell _{n,m,t}},\forall \,t\, \in {{\Omega }^T}
    \label{loss}
\end{equation}
\vspace{-5pt}
\subsubsection{LMO Model}
The LMO clears the LEM by considering the prosumers' bidding signals. In addition, the LMO participates in the WEM to buy or sell any required or excess energy. As a non-profit entity, the LMO aims to maximize social welfare while taking into account the WEM price. The LMO's objective is outlined in \eqref{lmo_ob}. The LMO considers the power balance in the LEM which is adhered by \eqref{lmo_bal}. 
\begin{equation}
\textbf{Min}\,\,\sum\limits_{t \in {{\Omega }^T}} {p_t^{\text{UG}}.\lambda _t^{\text{WEM}}.\Delta t}
    \label{lmo_ob}
\end{equation}
\vspace{-5pt}
\begin{equation}
p_t^{\text{UG}} = \sum\limits_{a \in {{\Omega }^A}} {p_{a,t}^{\text{net}}}  + p_t^{\text{Loss}},\forall \,t\, \in {{\Omega }^T}
    \label{lmo_bal}
\end{equation}
\subsubsection{Prosumer Model}
Prosumers participate in the LEM to maximize their profits. Alternatively, their resources can be aggregated by an aggregator agent. In such cases, aggregator agents collaborate with prosumers, aggregate their resources, and participate in the LEM based on the characteristics of the aggregated portfolio. The following presents a model for each prosumer ($a \in {{\Omega }^A}$) managing several DERs. This model is general and applicable to any prosumer with various resources, as well as similar agents such as aggregators and DG owners. The objective of a prosumer is as follows:
\begin{equation}
\textbf{Min} \sum\limits_{t \in {{\Omega }^T}} {\left[ \sum\limits_{n \in {{\Omega}^N_{a}}}{p_{a,t}^{\text{net}}.\,\lambda _{n,t}^{\text{LEM}}.\Delta t + \,\sum\limits_{v \in {\Omega }_a^V} {c_{v,t}^V} } \right]} \,
    \label{pros_obj}
\end{equation}
Where, $p_{{a,t}}^{\text{net}}$ is the prosumer's net consumption power (equation \eqref{pnet}). The prosumer's generation and consumption power are determined by \eqref{pg} and \eqref{pl}, respectively.
\begin{equation}
p_{{a,t}}^{\text{net}} = p_{a,t}^{\text{L}} - p_{a,t}^{\text{G}}, \forall \,\,t \in {{\Omega }^T}
    \label{pnet}
\end{equation}
\begin{equation}
p_{a,t}^{\text{G}} = \sum\limits_{v \in \left\{ {\text{PV}} \right\}} {p_{a,v,t}^V}  + \sum\limits_{v \in \text{\{ESS,EV\} }} {p_{a,v,t}^{\text{dCh-V}}} ,\forall \,\,t \in {{\Omega }^T}
    \label{pg}
\end{equation}
\begin{equation}
p_{a,t}^{\text{L}} = \sum\limits_{v \in \text{\{FLs\} }} {p_{a,v,t}^V}  + \sum\limits_{v \in \text{\{ESS,EV\} }} {p_{v,t}^{\text{Ch-V}}} ,\forall \,\,t \in {{\Omega }^T}
    \label{pl}
\end{equation}
%
The generated power from prosumers’ PVs is constrained by the minimum of forecasted generation capability and the capacity of the PV unit's inverter as described in \eqref{p_pv}. 
\begin{equation}
p_{a,u,t}^{\text{PV}} \le \textbf{Min}\left\{ {P_{a,u,t}^{\text{PV}},PF_{a,u,t}^{\text{PV}}.\overline S_{a,u}^{\text{PV}}} \right\},\forall u \in \Omega _a^{U},t \in {\Omega ^T}
    \label{p_pv}
\end{equation}
%
        \label{xch_dch_ess}
%
%
%
%
%
%
\vspace{-1mm}
Prosumers' EVs are modeled by \eqref{xch_dch_ev}-\eqref{soc_ev_l3}. The charging and discharging operations of EVs are represented by \eqref{xch_dch_ev}-\eqref{pch_ev}, while the EV's SoC is modeled by \eqref{soc_ev1}-\eqref{soc_ev_l3}. Constraint \eqref{soc_ev_l3} ensures that the EV's battery is sufficiently charged to provide the necessary energy for the prosumer's next trip. The BESSs can be modeled similarly to EVs, but unlike EVs, they can operate continuously throughout the day. 

\begin{equation}
x_{a,e,t}^{\text{Ch-EV}} + x_{a,e,t}^{\text{dCh-EV}} \le 1, \forall \,e \in \Omega _a^E,t \in \left[ {a_{a,e}^{\text{EV}},\,d_{a,e}^{\text{EV}}} \right]
\label{xch_dch_ev}
\end{equation}
\vspace{-5 pt}
\begin{equation}
x_{a,e,t}^{\text{Ch-EV}} = 0,\,\,\,\,x_{a,e,t}^{\text{dCh-EV}} = 0, \forall \,\,e \in \Omega _a^E,\,t \in {\Omega ^T} - \left[ {a_{a,e}^{\text{EV}},\,d_{a,e}^{\text{EV}}} \right]
        \label{xch_dch_ev2}
\end{equation}
\vspace{-5 pt}
\begin{equation}
0 \le p_{a,e,t}^{\text{Ch/dCh-EV}} \le x_{a,e,t}^{\text{Ch/dCh-EV}}\overline P_{a,e}^{\text{Ch/dCh-EV}}, \forall \,\,e \in \Omega _a^E,t \in {\Omega ^T}\label{pch_ev}
\end{equation}
\vspace{-5 pt}
\begin{equation}
\begin{array}{l}
SoC_{a,e,t}^{\text{EV}} = E_{a,e}^{\text{0-EV}} + \left( {\eta _{a,e}^{\text{Ch-EV}}p_{a,e,t}^{\text{Ch-EV}}\, - \frac{1}{{\eta _{a,e}^{\text{dCh-EV}}}}p_{a,e,t}^{\text{dCh-EV}}} \right)\Delta t,\;\\
\forall \,e \in {\Omega }_a^E,t = a_{a,e}^{EV}
\end{array}\label{soc_ev1}
\end{equation}
\vspace{-5 pt}
\begin{align}
\begin{small}
\begin{aligned}
SoC_{a,e,t}^{EV} &= SoC_{a,e,t-1}^{EV} + \left( \eta _{a,e}^{\text{Ch-EV}} p_{a,e,t}^{\text{Ch-EV}} - \frac{1}{\eta _{a,e}^{\text{dCh-EV}}} p_{a,e,t}^{\text{dCh-EV}} \right) \Delta t, \\
&\forall \, e \in \Omega_a^E, \, a_{a,e}^{EV} < t \le d_{a,e}^{EV}
\end{aligned}
\end{small}
\label{soc_ev2}
\end{align}
\vspace{-5 pt}
\begin{equation}
0 \le SoC_{a,e,t}^{EV} \le \overline {S}_{a,e}^{\text{EV}}, \forall \,\,e \in \Omega _a^E,t \in \left[ {a_{a,e}^{\text{EV}},\,d_{a,e}^{\text{EV}}} \right]
        \label{soc_ev_l1}
\end{equation}
\begin{equation}
SoC_{a,e,t = d_{a,e}^{\text{EV}}}^{\text{EV}} \ge E_{a,e}^{\text{trip-EV}}, \forall \,e \in {\Omega }_a^E
        \label{soc_ev_l3}
\end{equation}
%

The prosumers' FLs can be either interruptible or shiftable, modeled by \eqref{pil}-\eqref{eil}. Constraint \eqref{pil} ensures that the utilized flexibility remains within the allowable bounds, while \eqref{til} limits the number of changes in the load patterns. Finally, \eqref{eil} ensures that the prosumers' minimum required energy is supplied after the utilization of their FLs.
\vspace{-3mm}
\begin{equation}
-y_{a,t}^{\text{FL}}\overline P_{a,t}^{\text{FL}} \le p_{a,t}^{\text{FL}} \le y_{a,t}^{\text{FL}}\overline P_{a,t}^{\text{FL}}, \forall \,t \in {\Omega ^T}
    \label{pil}
\end{equation}
\vspace{-4mm}
\begin{equation}
\sum\limits_{t \in {\Omega ^T}} {y_{a,t}^{\text{FL}}}  \le \overline T_a^{\text{FL}}, \forall \,t \in {\Omega ^T}
        \label{til}
\end{equation}
\vspace{-4mm}
\begin{equation}
\sum\limits_{t \in {{\Omega }^T}} {{\bf{(}}P_{a,t}^{\text{L}} - p_{a,t}^{\text{FL}}{\bf{)}}} \,\Delta t \ge \underline E _a^{\text{L}}, \forall \,t \in {{\Omega }^T}
        \label{eil}
\end{equation}

\vspace{-4mm}
\subsection{Solution Method}\label{solution}
\vspace{-2mm}
To clear the LEM—achieving optimal energy pricing and the optimal strategy for each agent—a distributed framework based on a modified ADMM method is developed. In this framework, each agent (prosumers, DSO, and LMO) solves its optimization problem based on the received signals at each iteration and sends updated signals reflecting its strategy to the other agents. This method preserves the privacy of all agents and its convergence can be mathematically proven \cite{distri-jacquot2020dlmp}. \textcolor{black}{Unlike commonly used ADMM methods \cite{admm2-ullah2022dlmp} which require the exchange of agents' internal variables, the proposed ADMM approach incorporates a fictitious nodes’ injected power ($\tilde p_{a,t}^{\text{net}}$) and network power losses ($\tilde p_t^{\text{Loss}}
$) as auxiliary variables, to form the augmented Lagrangian function and establish connections between the agents’ problems.} This approach helps preserve agents’ privacy by avoiding the need to exchange internal characteristics among agents.

\begin{figure}
    \centering
    \includegraphics[width=0.43\textwidth]{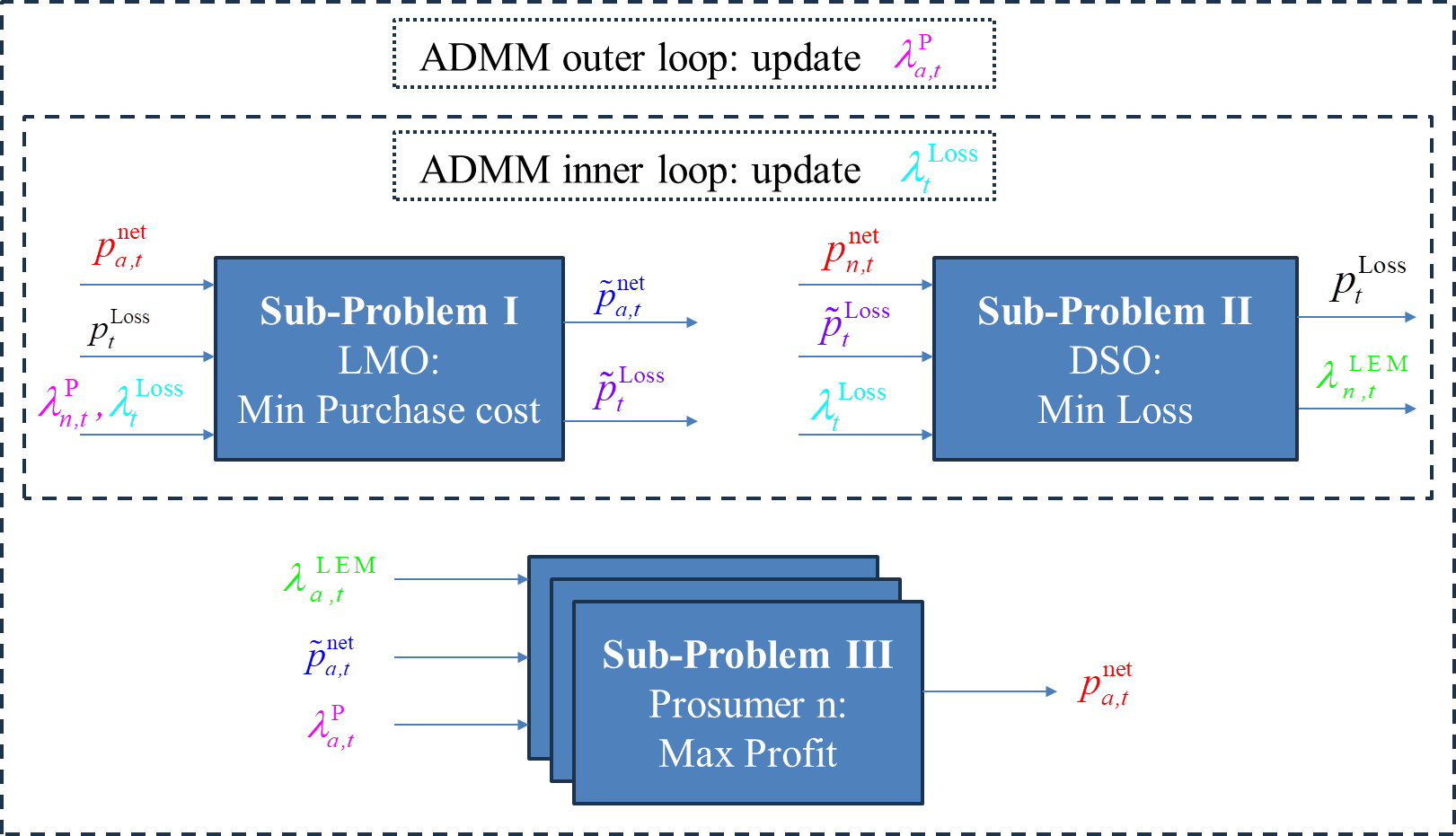}
    \vspace{-3mm}
    \caption{Sub-problems in two loops ADMM method.}
    \vspace{-4mm}
    \label{ADMM}
\end{figure}

The proposed solution method utilizes two ADMM loops: an outer loop and an inner loop. The outer loop manages the interaction between the LMO and prosumers, while the inner loop handles the interaction between the LMO and DSO. The augmented Lagrangian function derived from the centralized problem formulation is as follows, where ${\rho _a}$  and ${\rho '_a}$  are small positive numbers used in the ADMM method.
\vspace{-5 pt}
\begin{equation}
\begin{array}{l}
\mathcal{L}= \,\sum\limits_{t \in {{\Omega }^T}} {\left( {p_t^{\text{UG}}\lambda _t^{\text{WEM}} + p_t^{\text{Loss}}C_t^{\text{Loss}}} \right)\Delta t}  + \\
\sum\limits_{t \in {{\Omega }^T}} {\sum\limits_{a \in {{\Omega }^A}} {\left[ {\sum\limits_{n \in {\Omega }_a^N} {p_{a,t}^{\text{net}}.\,\lambda _{n,t}^{\text{LEM}}} \Delta t + \,\sum\limits_{v \in {\Omega }_a^V} {c_{v,t}^{V}} } \right]} } \\
 + \sum\limits_{t \in {{\Omega }^T}} {\sum\limits_{a \in {{\Omega }^A}} {\left( {\lambda _{a,t}^{\text{P}}\left( {\tilde p_{a,t}^{\text{net}} - p_{a,t}^{\text{net}}} \right) + \frac{{{\rho _a}}}{2}{{\left( {\tilde p_{a,t}^{\text{net}} - p_{a,t}^{\text{net}}} \right)}^2}} \right)} }  + \\
\sum\limits_{t \in {{\Omega }^T}} {\left( {\lambda _t^{\text{Loss}}\left( {\tilde p_t^{\text{Loss}} - p_t^{\text{Loss}}} \right) + \frac{{{{\rho '}_a}}}{2}{{\left( {\tilde p_t^{\text{Loss}} - p_t^{\text{Loss}}} \right)}^2}} \right)} \,
\end{array}
    \label{lagran}
\end{equation}
The following sub-problems are defined to clear the LEM. The sub-problems associated with each agent and the transaction signals between them are illustrated in Fig. \ref{ADMM}.
\subsubsection{Subproblem I (LMO problem)}
The LMO solves the following optimization problem:
\begin{equation}
\begin{array}{l}
\textbf{Min}\,\,\,\sum\limits_{t \in {{\Omega }^T}} {p_t^{\text{UG}}\lambda _t^{\text{WEM}}\Delta t}  +\\ \sum\limits_{t \in {{\Omega }^T}} {\sum\limits_{a \in {{\Omega }^A}} {\left(
\begin{array}{l}
{\lambda _{a,t}^{\text{P}}\left( {\tilde p_{a,t}^{\text{net}} - p_{a,t}^{\text{net(*)}}} \right) + \frac{{{\rho _a}}}{2}{{\left( {\tilde p_{a,t}^{\text{net}} - p_{a,t}^{\text{net(*)}}} \right)}^2}} \end{array} \right)}} \,\\
+\sum\limits_{t \in {{\Omega }^T}} {\left( {\lambda _t^{\text{Loss}}\left( {\tilde p_t^{\text{Loss}} - p_t^{\text{Loss(*)}}} \right) + \frac{{{{\rho '}_a}}}{2}{{\left( {\tilde p_t^{\text{Loss}} - p_t^{\text{Loss(*)}}} \right)}^2}} \right)} \, 
\end{array}
    \label{sp1}
\end{equation}

Subject to: \eqref{lmo_bal}\\
The problem is solved based on the received net consumption power of each prosumer, denoted by $p_{a,t}^{\text{net(*)}}$, and the DSO’s network losses, denoted by $p_t^{\text{Loss(*)}}$. (*) at top of each variable shows that it has specified amount from previous itterations. 
\subsubsection{Subproblem II (DSO problem)}
The DSO solves the following optimization problem:
\begin{equation}
\textbf{Min}\,\,\,\sum\limits_{t \in {{\Omega }^T}} {p_t^{\text{Loss}}.C_t^{\text{Loss}}.\Delta t}  + \sum\limits_{t \in {{\Omega }^T}} {\left( \begin{array}{l}
\lambda _t^{\text{Loss}}\left( {\tilde p_t^{\text{Loss}} - p_t^{\text{Loss}}} \right) + \\
\frac{{{{\rho '}_a}}}{2}{\left( {\tilde p_t^{\text{Loss}} - p_t^{\text{Loss}}} \right)^2}
\end{array} \right)} \,
    \label{BFM}
\end{equation}
Subject to: \eqref{active_bal} – \eqref{loss}.\\
The DSO determines the DLMP ($\lambda _{n,t}^{\text{LEM}}$) at each bus based on the signals received from the LMO, including the net consumption power at each bus and $\tilde p_t^{\text{Loss}}$. The LMO sends the net consumption power of each node to the DSO, which is derived from the prosumers’ net consumption power and their locations on the network, as detailed in equation \eqref{agg_nod}. 
\begin{equation}
p_{n,t}^{\text{net}} = \Psi _{n,a}.p_{a,t}^{\text{net}}
\label{agg_nod}
\end{equation}
The calculated DLMP will be sent to the corresponding prosumer based on its location in the network, as \eqref{agg_nod2}. 
\begin{equation}
\lambda _{a,t}^{\text{LEM}} = \Psi' _{n,a}.\lambda _{n,t}^{\text{LEM}}
    \label{agg_nod2}
\end{equation}
\subsubsection{Subproblem III (Each prosumer ($a \in {{\Omega }^A}$) problem)}
Each prosumer solves the following problem based on $\lambda _{a,t}^{\text{LEM (*)}}$ and $\tilde p_{a,t}^{\text{net (*)}}$, which are received from the LMO, to determine its optimal strategy (internal variables) and the net consumption power ($p_{a,t}^{\text{net}}$ ), which is then sent to the LMO.
\vspace{-5 pt}
\begin{equation}
\begin{array}{l}
\textbf{Min}\sum\limits_{t \in {{\Omega }^T}} {\left[ {p_{a,t}^{\text{net}}.\,\lambda _{a,t}^{\text{LEM(*)}}\Delta t - \,\sum\limits_{v \in {\Omega }_a^V} {c_{v,t}^{V}} } \right]}\\ 
+ \sum\limits_{t \in {{\Omega }^T}} {\left( \begin{array}{l}
\lambda _{a,t}^{\text{P}}\left( {\tilde p_{a,t}^{\text{net(*)}} - p_{a,t}^{\text{net}}} \right) +
\frac{{{\rho _a}}}{2}{\left( {\tilde p_{a,t}^{\text{net(*)}} - p_{a,t}^{\text{net}}} \right)^2}
\end{array} \right)} \,
\end{array}
    \label{sp3}
\end{equation}
Subject to: \eqref{pnet} - \eqref{eil}.\\     
In summary Algorithm \ref{alg} shows the LEM clearance approach.
\vspace{-15 pt}
\begingroup
\begin{algorithm}
\begin{algorithmic}[1]
\STATE \textbf{\textit{INIT}} required data: DSO, LMO, and prosumers parameters, stopping criteria, ${\rho _a}$, ${\rho '_a}$, $\lambda _{a,t}^{\text{P(1)}}$, $\lambda _t^{\text{Loss(1)}}$
\STATE $k = 1$, $k' = 1$ 
\STATE\textbf{\textit{While}} Stopping Criteria 1 is not true \textbf{\textit{do}} 
\STATE\hspace{2em}\textbf{\textit{For}} Each prosumer $a \in {\Omega^A}$ \textbf{\textit{do}}
\STATE\hspace{4em}Receive $\lambda _{a,t}^{\text{LEM (}k\text{)}}$, $\tilde p_{a,t}^{\text{net (}k\text{)}}$, $\lambda _{a,t}^{\text{P(}k\text{)}}$     
\STATE\hspace{4em}Solve Sub-problem III and attain: Internal variables and $p_{a,t}^{\text{net}\left( k \right)}$
\STATE\hspace{4em}Transmit $p_{a,t}^{\text{net}\left( k \right)}$ to LMO \STATE\hspace{2em}\textbf{\textit{End For}}
\STATE\hspace{2em}\textbf{\textit{While}} Stopping Criteria 2 is not true \textbf{\textit{do}}
\STATE\hspace{4em}DSO receives $p_{n,t}^{\text{net}\left( k \right)}$, $\tilde p_t^{\text{Loss (}k'\text{)}}$, $\lambda _t^{\text{Loss(}k'\text{)}}$
\STATE\hspace{4em}DSO solves Subproblem II and attains: Internal variables, $p_t^{\text{Loss-Grid}\left( {k'} \right)}$, and $\lambda _{n,t}^{\text{LEM}\left( {k'} \right)}$ 
\STATE\hspace{4em}DSO transmits $p_t^{\text{Loss-Grid}\left( {k'} \right)}$ to LMO
\STATE\hspace{4em}LMO Solves Sub-problem I for the received $p_{a,t}^{\text{net(}k\text{)}}$ from prosumers and $p_t^{\text{Loss-Grid}\left( {k'} \right)}$ from DSO 
\STATE\hspace{4em}LMO attains $\tilde p_t^{\text{Loss}\left( {k'} \right)}$ and send to the DSO
\STATE\hspace{4em}Update $\lambda _t^{\text{Loss}}$ as:\\
\hspace{4em}$\begin{array}{l}
\lambda _t^{\text{Loss}\left( {k' + 1} \right)} = \lambda _t^{\text{Loss}\left( {k'} \right)}\\\hspace{4em} + {\rho '_a}\left( {\tilde p_t^{\text{Loss}\left( {k'} \right)\,} - p_t^{\text{Loss-Grid}\left( {k'} \right)}} \right)\,\end{array}$
\hspace{4em} 
\STATE\hspace{4em}Update stopping criteria 2:\\
\hspace{4em}$\left| {\lambda _t^{\text{Loss}\left( {k'+1} \right)} - \lambda _t^{\text{Loss}\left( {k'} \right)}} \right| \le {\varepsilon _2}$
\STATE\hspace{4em}$k' = k' + 1$
\STATE\hspace{2em}\textbf{\textit{End While}}
\STATE\hspace{2em}LMO sends $\lambda _{a,t}^{\text{LEM}\left( k \right)}$ and $\tilde p_{a,t}^{\text{net}\left( k \right)}$  to the corresponding prosumers
\STATE\hspace{2em}Update $\lambda _{a,t}^{\text{P}}$ as:\\
\hspace{2em}$\lambda _{a,t}^{\text{P}\left( {k+1} \right)} = \lambda _{a,t}^{\text{P}\left( k \right)} + {\rho _a}\left( {\tilde p_{a,t}^{\text{net}\left( k \right)} - p_{a,t}^{\text{net}\left( k \right)\,}} \right)\,$
\STATE\hspace{2em}Update the stopping criteria 1:\\
\hspace{2em}$\left| {\lambda _{a,t}^{\text{P}\left( {k+1} \right)} - \lambda _{a,t}^{\text{P}\left( k \right)}} \right| \le {\varepsilon _1}$
\STATE\hspace{2em}$k=k+1$
\STATE\textbf{\textit{End While}}
\caption{LEM clearance algorithm.}
    \label{alg}
    \end{algorithmic}
\end{algorithm}
\endgroup
%
\section{Numerical Results}\label{results}
The proposed model is implemented on a modified IEEE 69-bus test system. The data of the network elements and peak load are introduced in \cite{IEEE-69-savier2007impact}. We assumed 30\% of customers under each load point are active prosumers. The active prosumers under each load node, are aggregated by an aggregator agent which participate in LEM to attain the optimal operation strategy of prosumers' resources. Each prosumer has a BESS, an EV, a PV unit, and FLs. The amount of prosumers' FLs is assumed to be 5\% of its normal load consumption at each time interval and at the end of the day, 95\% of total energy consumption should be adhered to. The load consumption pattern, WEM price and solar irradiation data are adopted from PJM market \cite{PJM-nyangon2023estimating}. We assume the PV units have identical output power due to similar geographical conditions. The average power factor of load nodes is equal to 0.85 lag. The BESSs and EVs technical parameters are adopted from \cite{pourghaderi2020energy}. Moreover, the data associated with EVs' arrival and departure, are generated by Monte-Carlo simulation technique using NHTS data \cite{NHTS-FHWA2022}.   

\begin{figure}
    \centering    \includegraphics[width=0.45\textwidth]{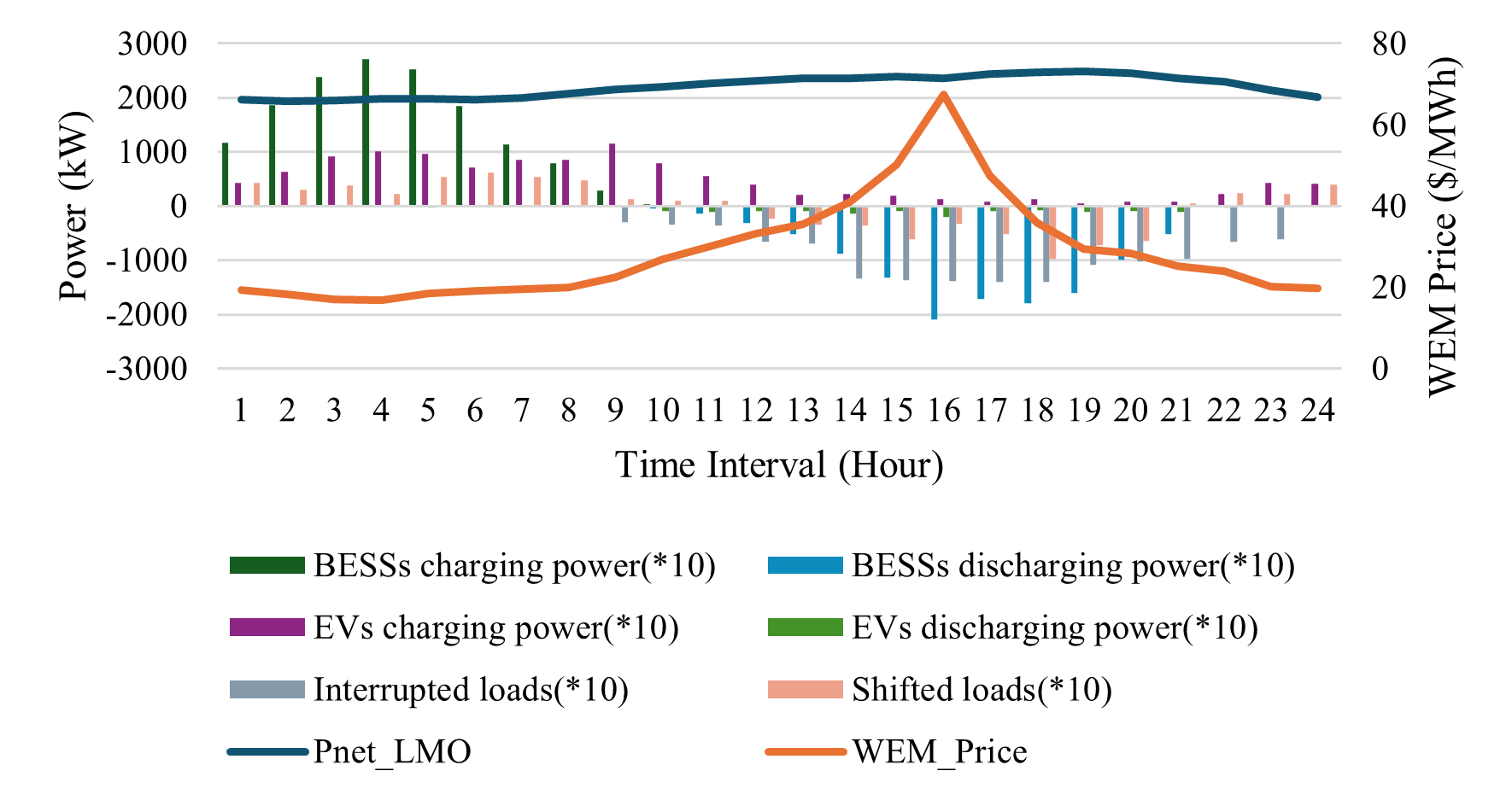}
    \vspace{-4mm}
    \caption{LMO's and prosumers' resources optimal strategy.}
    \vspace{-5mm}
    \label{LMO}
\end{figure}

\begin{figure}
    \centering
\includegraphics[width=0.45\textwidth]{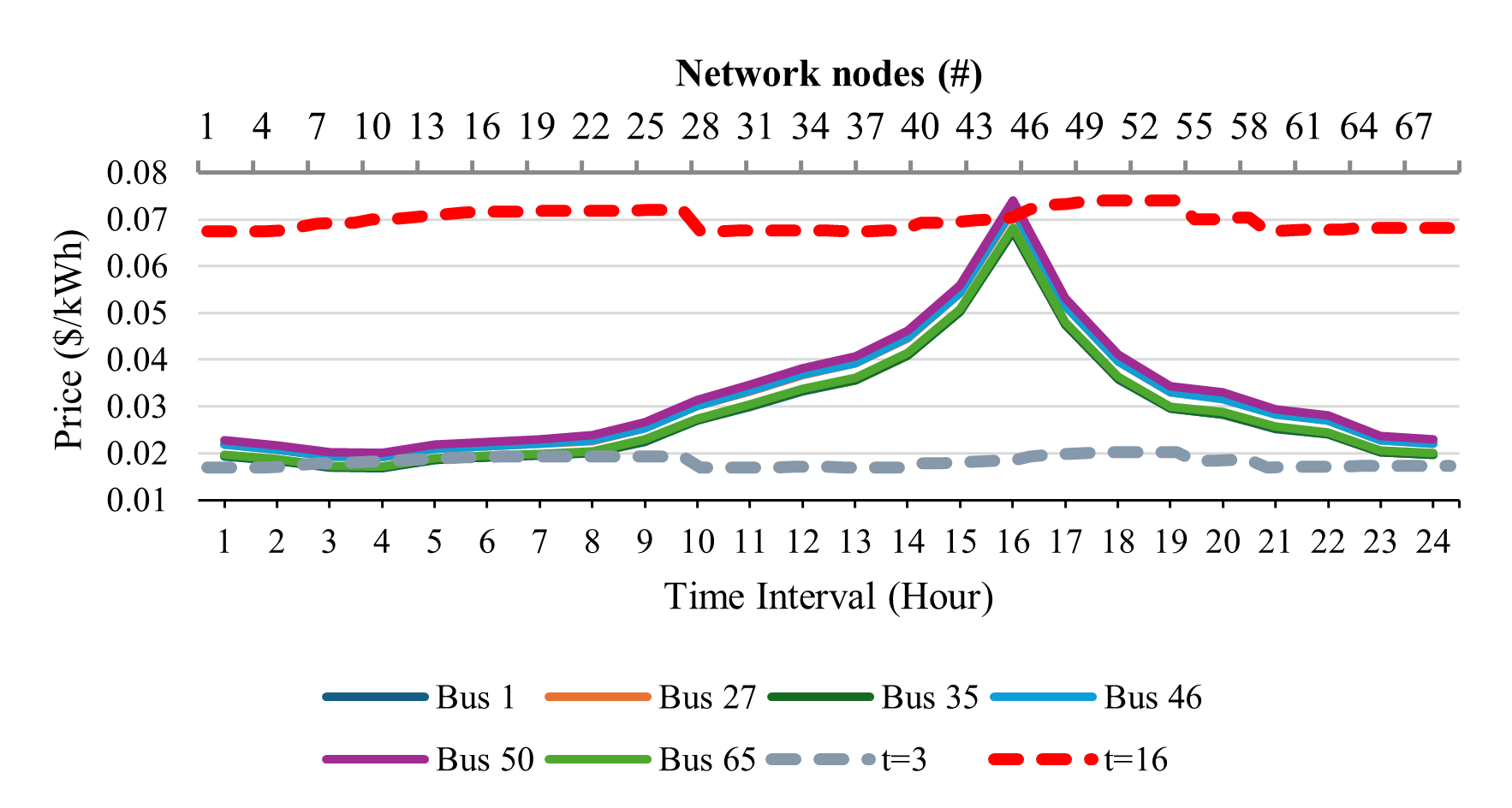}
    \vspace{-4mm}
    \caption{DLMP of substation and 5 end feeder nodes.}
    \vspace{-5mm}
    \label{DLMP}
\end{figure}

The LMO aims at maximizing the social welfare by minimizing the total purchase cost from WEM. Fig. \ref{LMO} shows the total purchased power from WEM. As it can be seen, the LMO clears the LEM in the manner to reduce its purchased power at the peak price period. It is done by adjusting LEM price through collaboration with DSO and prosumers.  
The aggregated charging and discharging profile of prosumers' BESSs and EVs are depicted in Fig. \ref{LMO}. The EVs charge/discharge schedule are optimized by the corresponding prosumer in which the associated prosumer's welfare constraints are considered. The optimal utilization of interruptible and defferable loads depicted in Fig. \ref{LMO}. The deffrable loads are utilized in the manner that the total consumed energy remains constant in the day while for the interrupted loads, the total consumed energy should be more than the minimum energy requirement of each prosumer. 
The DLMP at the substation bus and five ending bus of network feeders are compared in Fig. \ref{DLMP}. Moreover, in this figure the change in DLMP of two time intervals associated with minimum and maximum price with respect to different network nodes are illustrated. Regarding the convergence of the proposed two-loop ADMM method, the outer loop converges after 5 iterations, while the inner loop requires 3 iterations. The total CPU time is 22.27 seconds, making it an efficient approach for real-time applications.

To illustrate the effectiveness of the proposed model (Base Case), it is compared with three other cases. In Case I, each prosumer optimizes their strategy independently without collaborating with other agents in LEM. Case II uses a centralized approach, where a central agent (e.g., the DSO) minimizes the total energy purchase and loss costs without considering prosumers' preferences. In the last case (Case III), the framework is similar to the base case; however, the solution method relies on a mathematical approach used to solve bilevel problems, using the KKT condition and strong duality theorem \cite{Integ1-chen2021distribution}. The results are compared in Table \ref{Comparison}.

When comparing the proposed method with Case I, it can be observed that the DSO's energy losses and associated costs decrease by 2.6\% in the proposed method. Moreover, the total cost of the LMO in the WEM is reduced by 2.01\%, leading to an increase in social welfare. Notably, by implementing the proposed method, the LMO can, on average, decrease its costs by more than \$11,600 annually.
Comparing the proposed method with Case II, it is evident that while the associated costs of the LMO and DSO decrease, the costs incurred by prosumers increase. This is because the central agent disregards the prosumers' preferences. Regarding Case III, the results are almost the same as those of the base case since both have a similar framework. However, in Case III, the CPU time is significantly longer compared to the proposed method.
In summary, both Case II and Case III fail to preserve the agents' privacy, as all the internal data of prosumers and DSO's data is accessed by a central agent (in Case II) or a central solution machine (in Case III). Furthermore, the CPU time for these two cases is much longer than that of the proposed method, making them inefficient for real-time applications.

To clarify the positive impact of active prosumer participation in the LEM on the performance of the LMO, DSO, and the prosumers themselves, the effect of varying levels of active prosumer penetration \textcolor{black}{(i.e., the percentage of active prosumers relative to total number of customers) in the distribution network} is analyzed in Table \ref{prosumer_penetration}. The results indicate that a higher penetration of active prosumers leads to greater loss reduction for the DSO, while also reducing costs for the LMO. Furthermore, the total average costs for prosumers decrease. However, at higher penetration levels, the rate of cost reduction for all agents diminishes, indicating a saturation in prosumer penetration within the network.
\begin{table}[t]
    \centering
    \caption{Comparison of the proposed method with the case studies.}
    \vspace{-1mm}
    \resizebox{\columnwidth}{!}{%
        \begin{tabular}{@{}lcccc@{}}
            \toprule
            & \textbf{Base Case} & \textbf{Case I} & \textbf{Case II} & \textbf{Case III} \\ \midrule
            DSO’s costs (\$) & 141.24 & 145.01 & 137.24 & 141.29 \\
            LMO’s costs (\$) & 1550.12 & 1581.92 & 1548.14 & 1550.10 \\
            Avg. prosumers’ costs (\$) & 7.47 & 6.28 & 8.28 & 7.47 \\
            Data privacy & Yes & No & No & Yes \\
            CPU time & 22.27(s) & 1.36(s) & 2(h)-26(m)-32.2(s) & 12(h)-11(m)-12.4(s) \\ 
            \bottomrule
        \end{tabular}%
    }
\label{Comparison}
\end{table}
\begin{table}[t]
\centering
\caption{Comparison of agents' costs based on prosumers penetration.}
\vspace{-1mm}
\begin{tabular}{c|ccc}
\toprule
\multirow{2}{*}{\textbf{Prosumers penetration}} & \multicolumn{3}{c}{\textbf{Agents' cost in \$}} \\ \cmidrule(lr){2-4}
 & \textbf{LMO} & \textbf{DSO} & \textbf{Prosumers} \\ \midrule
45 & 1485.10 & 125.12 & 6.72 \\ \midrule
60 & 1432.23 & 112.32 & 6.02 \\ \midrule
75 & 1402.08 & 104.21 & 5.73 \\ 
\bottomrule
\end{tabular}
\label{prosumer_penetration}
\end{table}
\setlength{\textfloatsep}{0pt} 
\setlength{\floatsep}{0pt}
\setlength{\intextsep}{0pt}

\section{Conclusion} \label{conclusion}
This paper presents a fully distributed framework for clearing the LEM, taking into account the preferences of participating agents, including the LMO, DSO, and prosumers. The framework is based on a two-loop modified ADMM method, where the outer loop establishes interactions between the LMO and prosumers, while the inner loop addresses the coordination between the LMO and DSO. The proposed method effectively addresses the objectives and techno-economic constraints of all agents. Implementing the proposed method in the test system and comparing it with three case studies reveals its advantages. The framework preserves the agents' privacy, as only the prosumers' net power and the DLMP, which describes LEM clearing price, are exchanged between them. Moreover, the proposed algorithm converges to the optimal solution in less than a minute, making it an efficient approach for real-time applications.
\vspace{-2mm}
\section*{Acknowledgments}
\vspace{-2mm}
This publication is based upon work supported by King Abdullah University of Science and Technology under Award No. RFS-OFP2023-5505.
\vspace{-2mm}
\bibliographystyle{IEEEtran}
\bibliography{Refs}

\end{document}